\documentclass[manuscript]{revtex4-1}
\usepackage{amsmath,amssymb,graphicx,subfig,upgreek,siunitx}
\usepackage{graphicx}
\usepackage{color}

\begin{document}

\title{VO$_2$-based radiative thermal transistor in the static regime}
\author{Hugo Prod'Homme, Jose Ordonez-Miranda, Youn\`es Ezzahri, J\'er\'emie Dr\'evillon and Karl Joulain}
\affiliation{Institut Pprime, CNRS, Universit\'e de Poitiers, ISAE-ENSMA, F-86962 Futuroscope Chasseneuil, France}

\date{\today}
\begin{abstract}
We study a near-field radiative thermal transistor analogous to an electronic one made of a VO$_2$ base placed between two silica semi-infinite plates playing the roles of the transistor collector and emitter. The fact that VO$_2$ exhibits an insulator to metal transition is exploited to modulate and/or amplify heat fluxes between the emitter and the collector, by applying a thermal current on the VO$_2$ base. We study the transistor behavior in 4 typical regimes where the emitter-base and base-collector separation distances can be larger or smaller than the thermal wavelength, and in which the VO$_2$ layer can be opaque or transparent. Thermal currents variations with the base temperatures are calculated and analyzed. An optimum configuration for base thickness and separation distance maximizing the thermal transistor modulation factor is found.
\end{abstract}

\maketitle

\section{Introduction.}
Recent years have seen the increasing number of researches about thermal rectification e.g. the ability for a system to present a different heat flux when a temperature difference is imposed at the extremities of the system in a certain direction or reversed. Thermal rectification has lead to the development of thermal components such as thermal diodes \cite{li_thermal_2004,ben-abdallah_phase-change_2013,ito_experimental_2014,nefzaoui_simple_2014,nefzaoui_radiative_2014,ordonez-miranda_photonic_2017}, thermal transistor \cite{chung_lo_thermal_2008,ben-abdallah_near-field_2014,joulain_quantum_2016} and even thermal memories \cite{kubytskyi_radiative_2014} that could eventually give birth to the emergence of a new field named thermotronics.

Heat control researches could have important applications in energetics. Indeed, these thermal components could be used in a similar way as their electronic counterparts such as the electronic diode and the bipolar transistor, which have revolutionized the world during the 20th century. Electronic components guide, amplify and modulate electric currents. At present time, these functions hardly exist for thermal currents. However, the increasing energy demand, the limited energy resources as well as the global warming issues require a better energy management, and in particular heat control. Due to the limitation of Carnot efficiency, huge amounts of heat that is produced in energy processing is lost. This waste heat could be harvested if the thermal diode and the thermal transistor were reliably developed.
The goal of this paper is to exhibit the performances of this latter thermal component. Following the pioneer work of Ben-Abdallah and Biehs \cite{ben-abdallah_near-field_2014}, we study a radiative thermal transistor made of two plates of silica between which is placed a vanadium dioxide (VO$_2$) film. VO$_2$ is a material of very strong interest in thermal sciences and in particular thermal radiation due to its insulator to metal transition. Below a critical temperature of 341 K, VO$_2$ is a dielectric, whereas it is a metal above 346 K. Therefore, its optical properties dramatically change in a narrow temperature range and allow strong variations in the different heat fluxes exchanged between the plates.
In the devices presented here, the two pieces of silica are maintained at two different temperatures, while a thermal flux is applied to the VO$_2$ film. We study how the heat flux between the silica pieces is controlled by the thermal current applied to the VO$_2$ film for both the near and far field as well as for opaque of transparent VO$_2$ films. We show that a transistor effect can be obtained in all the regimes studied but with various performances. We find that a configuration for the base thickness and the separation distance between the base and the transistor therminals for optimizing the thermal modulation. These considerations could guide the future design and fabrication of thermal transistors.

\section{Principle of the radiative thermal transistor.}
\begin{figure}
\begin{center}
\includegraphics[width=15cm]{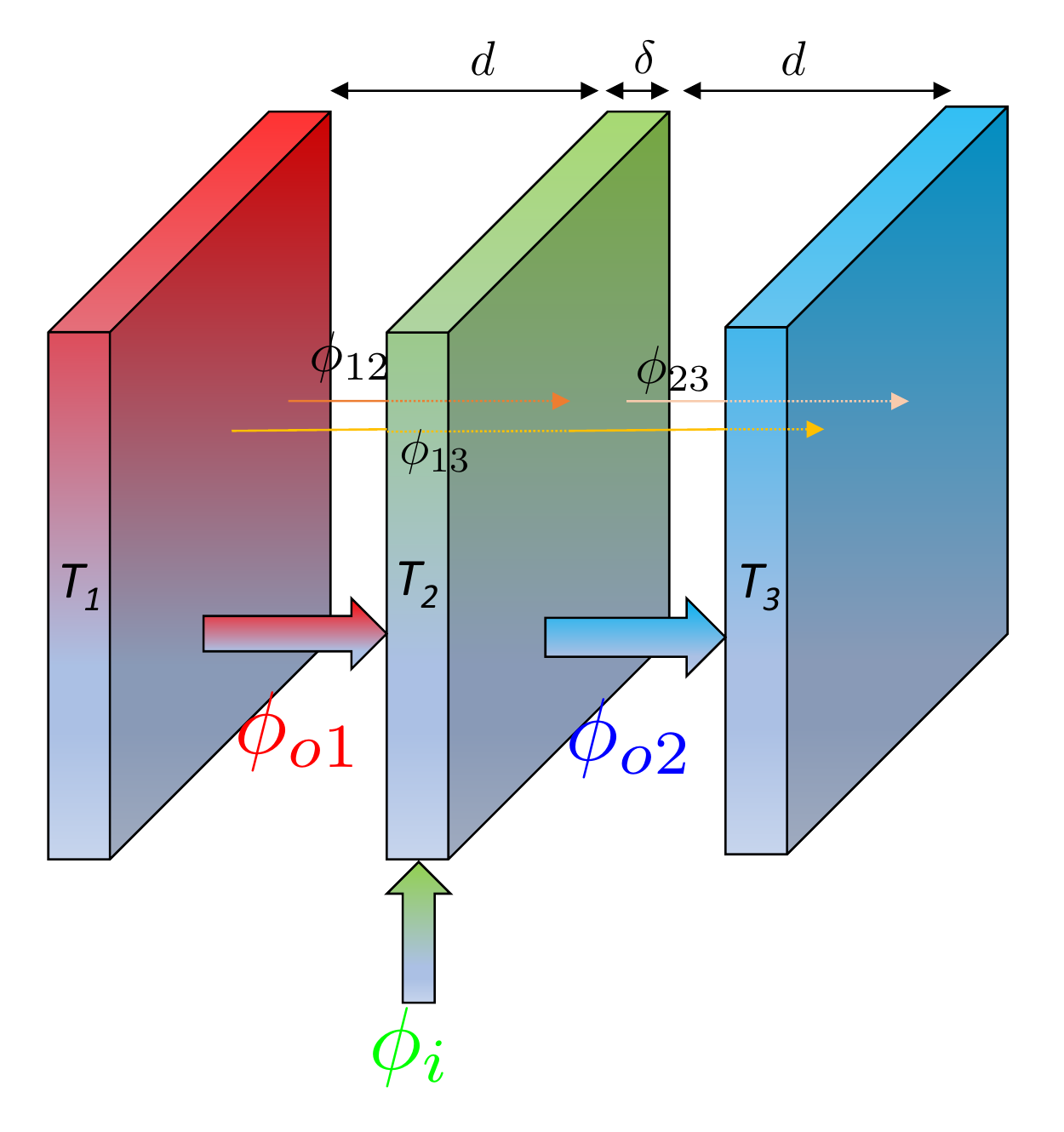}
\caption{Radiative thermal transistor constituted of 2 plates of SiO$_2$ (emitter and collector) and a VO$_2$ layer (base).}
\label{Transistor_Scheme}
\end{center}
\end{figure}

The device under consideration is constituted of 2 semi-infinite pieces of silica (SiO$_2$) between which is placed a VO$_2$ film (Fig. \ref{Transistor_Scheme}). The two semi infinite SiO$_2$ plates are maintained at temperatures $T_1$ or $T_3$. To do so, a thermal flux $\phi_{o1}$ and $\phi_{o2}$ has to be applied at these two materials since these two output fluxes are also the fluxes that are lost by the plates 1 and 3 . A third thermal current $\phi_i$ is also applied to the VO$_2$ layer which is placed between the two silica plates. The equilibrium temperature of the VO$_2$ plate is $T_2$. We call $d$ the distance between the plates and the films (which are equal) and we call $\delta$ the layer thickness. We solve the problem at thermal equilibrium so that $\phi_i=\phi_{o2}-\phi_{o1}$.
In analogy with an electronic bipolar transistor \cite{joulain_modulation_2015}, the two silica plates can be seen as the equivalent of an emitter and a collector whereas the VO$_2$ layer can be seen as the equivalent of a base. In our study, the three fluxes $\phi_{o1}$, $\phi_{o2}$ and $\phi_3$ will be plotted against the base temperature $T_2$. We will see that the base temperature $T_2$ can be changed by a little amount of heat $\phi_i$ compared to the emitter flux $\phi_{o1}$ and the collector flux $\phi_{o2}$.
The modulation efficiency at the collector and the emitter quantifies the device faculty to modulate heat currents. It is defined as $ME_l=\phi_{ol}^\mathrm{min}/\phi_{ol}^\mathrm{max}$ where $l=1,2$. 
The ability to amplify the current is quantified by a differential amplification factor $\alpha$ which is defined as the derivative of the emitter or collector current with respect to base current $\alpha_l=|\partial\phi_{ol}/\partial\phi_i|$. This coefficient therefore measures by how much an elementary current variation on the base is amplified on the collector-base or base-emitter heat current. The transistor regime is achieved when the amplification factor $\alpha$ is higher than 1.

\subsection{Calculation of the heat fluxes}
The heat transfer problem is more complicated than the standard ones that have been studied in the past \cite{ben-abdallah_near-field_2014,joulain_modulation_2015}. Indeed, in the present situation, the base is not necessarly opaque so that a thermal flux can be directly transmitted from the emitter to the collector ($\phi_{13}$). Output and input fluxes can be expressed from the individual heat fluxes exchanged between individuals bodies
\begin{eqnarray}
\phi_{o1} & = & \phi_{12}+\phi_{13} \\
\phi_{o2} & = &  \phi_{23}+\phi_{13}\\
\phi_i & = & \phi_{o2}-\phi_{o1}
\end{eqnarray}
These heat fluxes can be calculated using fluctuating electrodynamics formalism \cite{Rytov:1989ur,Polder:1971uu,joulain_surface_2005,Volokitin:2007el,bimonte_nonequilibrium_2017}. This formalism has been developped in several geometries like planar films \cite{francoeur_near-field_2008}, spheres \cite{narayanaswamy_thermal_2008,kruger_trace_2012} and even in arbitrary geometries by means of numerical tools \cite{rodriguez_fluctuating-surface-current_2012,polimeridis_fluctuating_2015}. The case of the three body configuration like it is the case here has been adressed by Messinal et al. \cite{messina_three-body_2012,messina_three-body_2014}. In this situation, the fluxes read 
\begin{equation}
\label{ }
\phi_{ij}=\sum_{\mathrm{pol}=s,p}\int\frac{d\omega}{2\pi}[\Theta(\omega,T_i)-\Theta(\omega,T_j)]\int \frac{KdK}{2\pi}\mathcal{T}_{i,j}^\mathrm{pol}(\omega,K)
\end{equation}
where
$$
\Theta(\omega,T)=\frac{\hbar\omega}{\exp[\hbar\omega/k_BT]-1}
$$
is the mean energy of a photon at angular frequency $\omega$  and at thermal equilibrium at temperature $T$. $K$ is the parrallel wavevector to the interface. The superscript pol denotes the polarization that can be $s$ or $p$. $\mathcal{T}_{i,j}^\mathrm{pol}(\omega,K)$ is the transmission factor of the mode $(K,\omega)$ between bodies $i$ and $j$ for the polarization pol. Between bodies 1 and 3, this transmission factor reads
\begin{equation}
\label{ }
\mathcal{T}_{1,3}^\mathrm{pol}(\omega,K)=X_1\left|\frac{\tau_2^\mathrm{pol}e^{2ik_zd}}{(1-\rho_1^\mathrm{pol}\rho_{23}^\mathrm{pol}e^{-ik_zd})(1-\rho_2^\mathrm{pol}\rho_3^\mathrm{pol}e^{-ik_zd})}\right|^2X_3
\end{equation}
where $k_z=\sqrt{\omega^2/c^2-K^2}$ and where the $X$ coefficients read
\begin{equation}
X_i=\begin{cases}
1-\left|\rho_i^\mathrm{pol}\right|^{2}-\left|\tau_i^\mathrm{pol}\right|^{2} & ,K<k_{0}\\
2\mathrm{Im}\left(\rho_i^\mathrm{pol}\right) & ,K>k_{0}
\end{cases}
\label{coeffs_X}
\end{equation}
In these expressions $\rho_i^\mathrm{pol}$ and $\tau_i^\mathrm{pol}$ are the Fresnel reflexion and transmission coefficient between vacuum and the layer of material $i$ of finite or infinite thickness. The subscript $i$ can be replaced by another subscript $ij$ to take into account an ensemble of two layers instead of a simple layer. In order to calculate the transmission coefficients between all the bodies, one has to introduce the transmission coefficient between one body and the ensemble of two bodies as
\begin{subequations}
\begin{align}
\mathcal{T}_{1,23}^\mathrm{pol}=X_{1}^\mathrm{pol}\left|\frac{e^{-ik_{z}d}}{1-\rho_{1}^\mathrm{pol}\rho_{23}^\mathrm{pol}e^{-2ik_{z}d}}\right|^{2}X_{23}^\mathrm{pol} \, , \\
\mathcal{T}_{12,3}^\mathrm{pol}=X_{12}^\mathrm{pol}\left|\frac{e^{-ik_{z}d}}{1-\rho_{12}^\mathrm{pol}\rho_{3}^\mathrm{pol}e^{-2ik_{z}d}}\right|X_{3}^\mathrm{pol} \, ,
\end{align}
\end{subequations}

The probability for a photon to go from body 1 to body 2 is equal to the probability to go from body 1 to the equivalent of bodies 2 and 3 \emph{minus} the probability to go from body 1 to body 3 \emph{and beyond} denoted $1_\infty$ in the following. This means that the transmission coefficient $\tau_3$ must be set equal to zero within the expression of $X_3$ for \eqref{eq:subtransmission1}. And similarly for $\tau_1$ regarding \eqref{eq:subtransmission2}.
As bodies 1 and 3 are of infinite lengths, their transmission functions are equal to zero.
\begin{subequations}
\begin{align}
\mathcal{T}_{1,2}^\mathrm{pol}=\mathcal{T}_{1,23}^\mathrm{pol}-\mathcal{T}_{1,3_{\infty}}^\mathrm{pol}
\label{eq:subtransmission1} \, , \\
\mathcal{T}_{2,3}^\mathrm{pol}=\mathcal{T}_{12,3}^\mathrm{pol}-\mathcal{T}_{1_{\infty},3}^\mathrm{pol}
\label{eq:subtransmission2} \, .
\end{align}
\label{eq:subtransmission}
\end{subequations}
The values of the fluxes $\phi_{12}$, $\phi_{23}$ and $\phi_{13}$ are obtained by computing the previously described expressions. Then, it is possible to define exchange functions $\mathcal{E}$ from the values of $\phi_{12}$, $\phi_{23}$ and $\phi_{13}$, that can be seen as normalized flux to the Stefan-Boltzmann law :
\begin{subequations}
\begin{align}
\phi_{12}=\sigma\mathcal{E}_{1}\left(T_{1},T_{2}\right)\left(T_{1}^{4}-T_{2}^{4}\right) \, , \\
\phi_{23}=\sigma\mathcal{E}_{2}\left(T_{2},T_{3}\right)\left(T_{2}^{4}-T_{3}^{4}\right) \, , \\
\phi_{13}=\sigma\mathcal{E}_{\mathrm{t}}\left(T_{1},T_{2},T_{3}\right)\left(T_{1}^{4}-T_{3}^{4}\right) \, .
\end{align}
\end{subequations}
When an exchange function between two bodies is equal to 1, the heat flux density between the bodies is equal to the heat transfer between two blackbodies.

We now describe the temperatures of the problem with new variables, considering that $T_1$ and $T_3$ are distributed upper and lower than a central temperature $T_{e}$, with a difference (which relates then to a temperature gradient) $T_\Delta$, in a power-4-temperature scale, such as described by the following relations :
\begin{subequations}
\begin{align}
T_{e}^4=\frac{T_1^4+T_3^4}{2} \, , \\
T_\Delta^4=\frac{T_1^4-T_3^4}{2} \, .
\end{align}
\end{subequations}
Note that $T_e$ corresponds to the equilibrium temperature of a blackbody situated between two blackbodies at temperature $T_1$ and $T_2$.
In terms of the new variables $T_e$ and $T_\Delta$, the output and input fluxes now read :
\begin{subequations}
\begin{align}
\phi_{{o1}}=2\sigma\left(\mathcal{\mathcal{E}}_{{e}}+\mathcal{\mathcal{E}}_{\mathrm{t}}\right)T_{\Delta}^{4}-\frac{\mathcal{\mathcal{E}}_{1}}{\mathcal{\mathcal{E}}_{1}+\mathcal{\mathcal{E}}_{2}}\phi_{{i}}
\label{eq:phio1ex} \, , \\
\phi_{\mathrm{o2}}=2\sigma\left(\mathcal{\mathcal{E}}_{{e}}+\mathcal{\mathcal{E}}_{{t}}\right)T_{\Delta}^{4}+\frac{\mathcal{\mathcal{E}}_{2}}{\mathcal{\mathcal{E}}_{1}+\mathcal{\mathcal{E}}_{2}}\phi_{{i}}
\label{eq:phio2ex} \, , \\
\phi_{{i}}=\sigma\left(\mathcal{E}_{1}+\mathcal{E}_{2}\right)\left(T_{2}^{4}-T_{{e}}^{4}\right)+\sigma\left(\mathcal{E}_{2}-\mathcal{E}_{1}\right)T_{\Delta}^{4}
\label{eq:phiiex} \, ,
\end{align}
\end{subequations}
with :
\begin{equation}
\mathcal{E}_{{e}}=\frac{\mathcal{E}_{1}\mathcal{E}_{2}}{\mathcal{E}_{1}+\mathcal{E}_{2}}
\, .
\end{equation}
Let us not interpret the variations of the fluxes by examining expression \eqref{eq:phio1ex}, \eqref{eq:phio2ex} and \eqref{eq:phiiex}. The first terms of the expressions \eqref{eq:phio1ex} and \eqref{eq:phio2ex} are identical and are actually the ones that relates to the transistor effect. The phase transition will lead to an important change in the values of $\mathcal{E}_\mathrm{e}+\mathcal{E}_\mathrm{t}$, so the output fluxes will be modulated by the variation of this factor. The factor $T_\Delta^4$ also shows that the maximum possible variation scales with the temperature difference, which is intuitively verified.

The second terms of both \eqref{eq:phio1ex} and \eqref{eq:phio2ex} have opposite signs. Since we consider that the base is equidistant from the collector and the emitter, we find that $\mathcal{E}_{1}\sim\mathcal{E}_{2}$, so that these terms can be approximated to $\pm\frac{1}{2}\phi_\mathrm{i}$. These terms express the fact that the input flux which is brought to the base must be redistributed to the collector or the emitter, following the conservation of the fluxes. 

The first term of \eqref{eq:phiiex} describes the fact that the more $T_2$ differs from $T_\mathrm{e}$, and the more $\phi_\mathrm{i}$ must be great to keep the temperature of the base constant thus to stay at thermal equilibrium. If $\mathcal{E}_{1}=\mathcal{E}_{2}$, the second term of \eqref{eq:phiiex} is null, therefore the base temperature $T_2$ will be equal to $T_\mathrm{e}$ in the absence of input flux ($\phi_\mathrm{i}=0$).
When the values of $T_\Delta$ become larger, the Planck energy distributions of the emitter $\Theta\left(\omega,T_1\right)$ and the collector $\Theta\left(\omega,T_3\right)$ are different so that $\mathcal{E}_{2}-\mathcal{E}_{1}$ will vary according to $T_2$. In this case the second term of \eqref{eq:phiiex} is not negligible and has an impact on the thermal behavior of the transistor. 

The advantage of this formalism is that the thermal behavior of the device is fully described by the exchanged functions $\mathcal{E}_1$, $\mathcal{E}_2$ and $\mathcal{E}_\mathrm{t}$ that are depending on $T_\mathrm{e}$, $T_\Delta$, $T_2$, $d$ and $\delta$.

Subsequently, the amplification factors are written :
\begin{eqnarray}
\alpha_1&=&2\sigma\frac{\mathrm{d}\mathcal{E}_{{e}}+\mathrm{d}\mathcal{E}_{{t}}}{\mathrm{d}\phi_{{i}}}T_{\Delta}^{4}-\frac{\mathcal{E}_1}{\mathcal{E}_1+\mathcal{E}_2}
 \nonumber\\
\alpha_2&=&2\sigma\frac{\mathrm{d}\mathcal{E}_{{e}}+\mathrm{d}\mathcal{E}_{{t}}}{\mathrm{d}\phi_{{i}}}T_{\Delta}^{4}+\frac{\mathcal{E}_2}{\mathcal{E}_1+\mathcal{E}_2} \label{eq:amplification}
\end{eqnarray}

After showing that this formalism is well-adapted to describe the transistor behavior in a simple case first, we will describe the variations of these exchange functions according to the different parameters in the following section.

\section{Results}
\begin{figure}
\includegraphics[width=\textwidth]{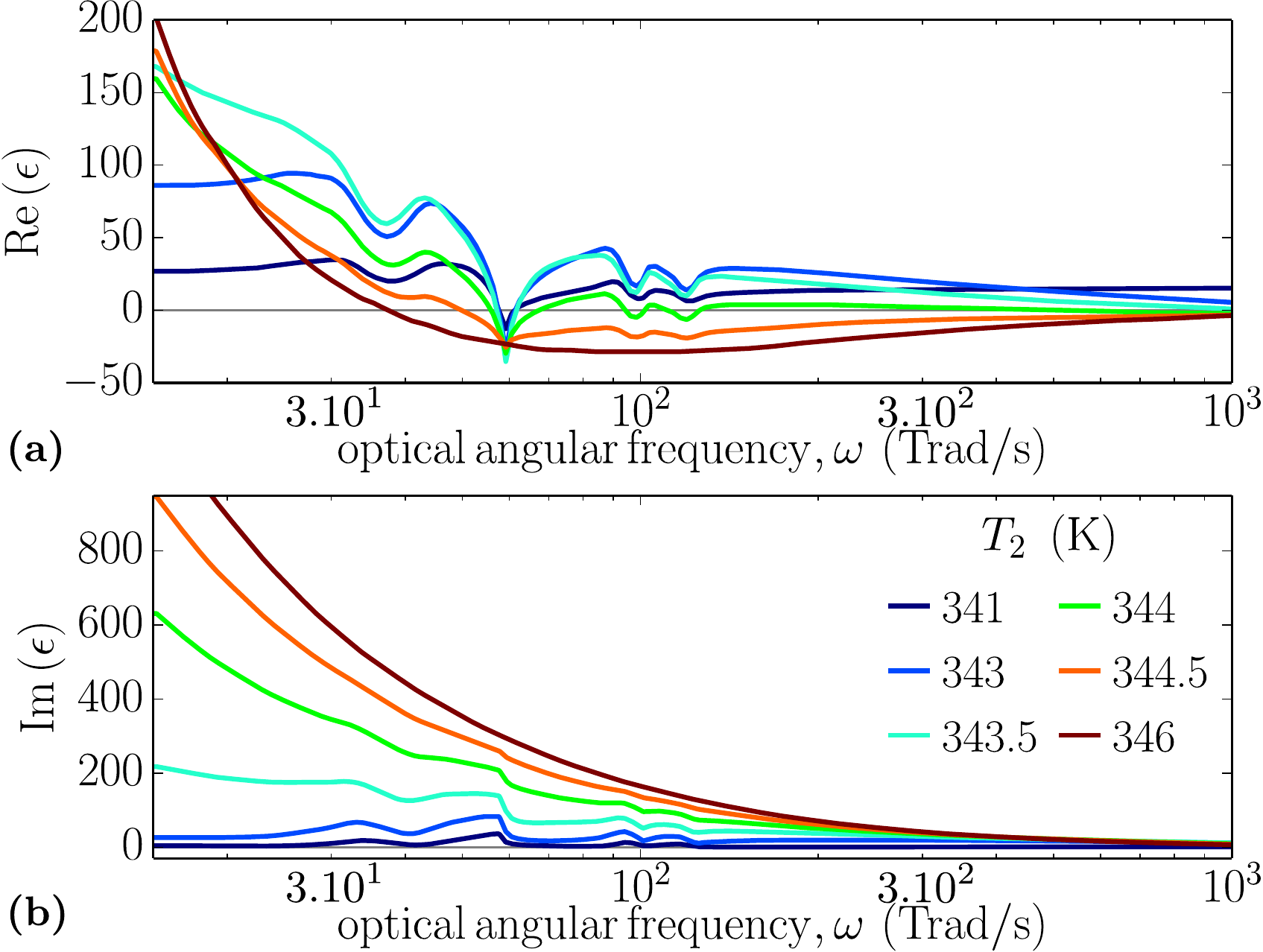}
\caption{(a) Real and (b) imaginary parts of $\mathrm{VO_2}$ permittivity variations with the angular frequency $\omega$, for different temperatures in the phase transition zone. The legend of (b) applies for (a) as well.\label{fig:permittivity}}
\end{figure}

\begin{figure}
\begin{center}
\includegraphics[width=8cm]{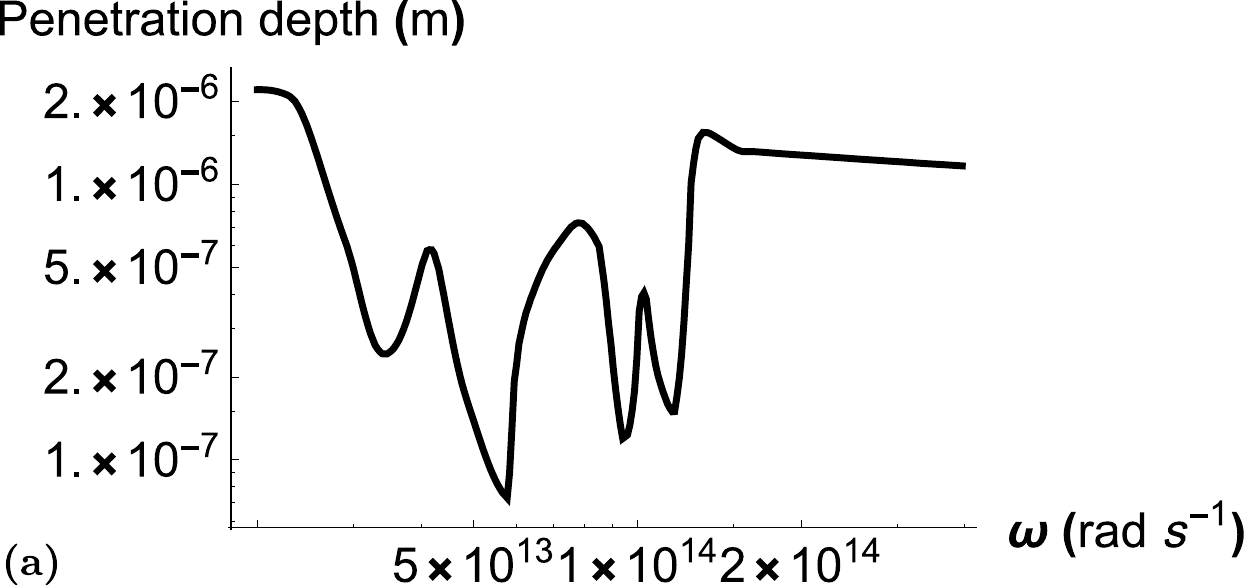}
\includegraphics[width=8cm]{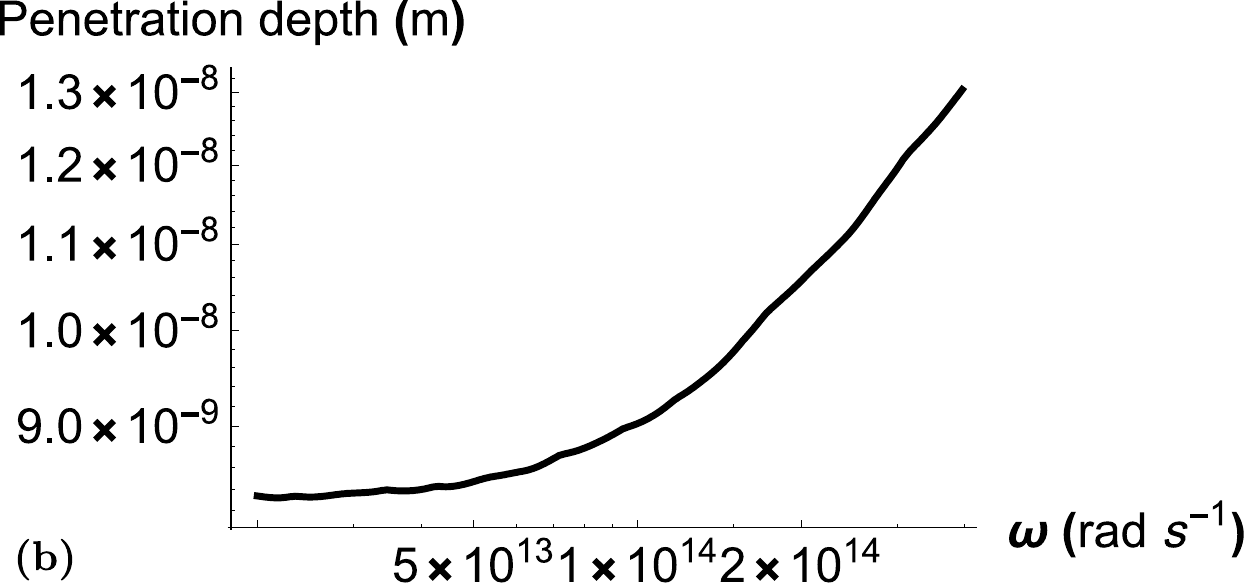}
\caption{Penetration depth $\delta_\mathrm{pen}$ of a plane wave in VO$_2$ in the dielectric phase (a) and the metallic phase (b).}
\label{Pendepth}
\end{center}
\end{figure}

Calculation of the output and input fluxes necessitates to know the material permittivities variation with the angular frequencies and with the temperature. Concerning VO$_2$ these permittivities are taken from \cite{qazilbash_mott_2007} whereas they are taken from \cite{joulain_radiative_2015} concerning SiO$_2$. Among these two materials, VO$_2$ is the one that exhibits the most drastic changes when the temperature is in the transition zone. Its variations are depicted in Fig. \ref{fig:permittivity}.
With these datas, we were able to calculate the output and input heat fluxes for any temperature at the collector, emitter and base.  We will now examine different cases in terms of the separation distance $d$ and the base thickness $\delta$. We will distinguish the near-field region where $d$ is much smaller than the thermal wavelength from the far-field region where $d$ is much larger than the thermal wavelength. At temperature between 300 K and 400 K the thermal wavelength is typically between 7.5 $\si{\um}$ and 10 $\si{\um}$. Two other regime will be of interest whether the base is transparent or opaque depending on the thickness $\delta$. In Fig. \ref{Pendepth}, we plot the penetration depth $\delta_\mathrm{pen}$ of a plane wave in VO$_2$ given by $\delta_\mathrm{pen}=c/[2\omega\mathrm{Im}(\epsilon(\omega))]$.
It can be seen that in the metallic phase, the penetration depth is very small so that the base is opaque when VO$_2$ is in this phase except when $\delta$ is lower than a few $\si{\um}$. However, in the dielectric phase, the base is opaque only if the base is larger than around ten microns and is semi-transparent as soon as $\delta$ is smaller than a few $\si{\um}$.

\subsection{Far-field transfer and opaque base.}
\begin{figure}
\includegraphics[width=\textwidth]{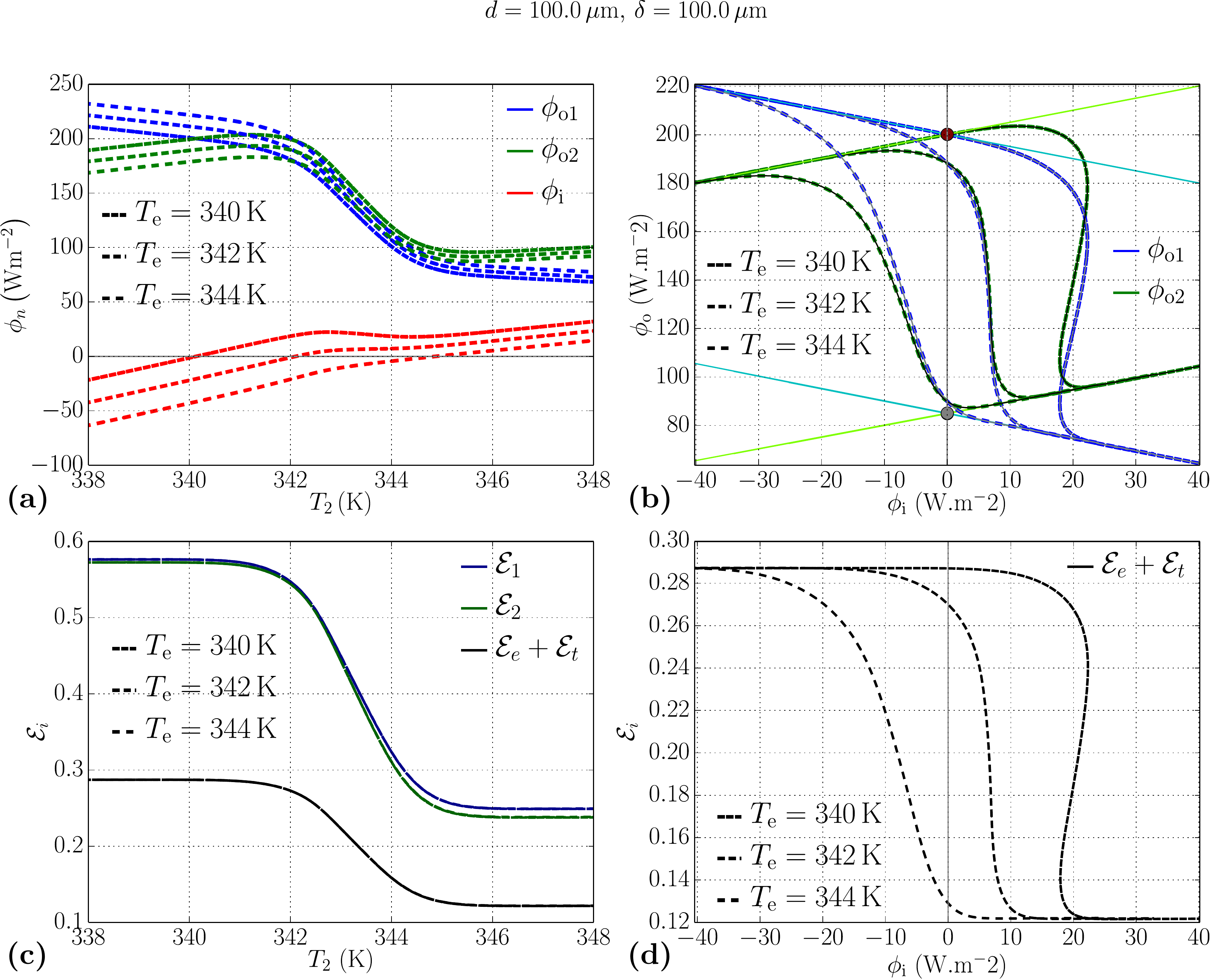}
\captionsetup{justification=centerlast}
\caption{Input and output radiative heat flux densities within the thermal transistor in the opaque ($\delta=100\,\si{\um}$) and far-field ($d=100\,\si{\um}$) cases, for three values of the central temperature ($T_{e}=340,342,344$ K) and $T_\Delta=280\,\mathrm{K}$.
(a) : ($\phi_{i}$) and output ($\phi_{o1},\phi_{o2}$) fluxes versus the base temperature $T_2$.
(b) : Output fluxes ($\phi_{o1},\phi_{o2}$) versus input flux $\phi_i$.
(c) : Exchange functions ($\mathcal{E}_1$, $\mathcal{E}_2$, $\mathcal{E}_e+\mathcal{E}_t$) versus $T_2$.
(d) : Exchange function $\mathcal{E}_e+\mathcal{E}_t$ versus $\phi_i$.}
\label{fig:simplecase}
\end{figure}

We now consider the case where the distance $d$ is large compares to the thermal wavelength and $\delta$ is large compared to the penetration depth in the VO$_2$. Let us take as an example $d=\delta=100$ $\si{\um}$.
First, we set $T_\Delta$ to 280 K, and we compute $\phi_{o1}$ and $\phi_{o2}$ according to $\phi_i$ for three values of $T_e$ : 340 K, 342 K and 344 K. (The respective values of $T_1$ are around 373.7 K, 375.2 K and 376.8 K and the respective values of $T_3$ are around 291.5 K, 294.6 K and 297.7 K.)
The values of the input and output fluxes as well as exchange functions are represented  versus $T_2$ and $\phi_i$ in Figure \ref{fig:simplecase}, for the three values of $T_{e}$. 

In the upper left plot of Fig. \ref{fig:simplecase}, the fluxes are plotted versus the base temperature $T_2$. We see that for the 3 values of $T_e$ considered, there is a value of $T_2$ where the input flux is 0. For this temperature, the base is at thermal equilibrium with the emitter and the collector. In this plot, we note that a small change in the variations of the flux $\phi_i$ will result in an important change in the temperature $T_2$ and in the output fluxes : this is the thermal transistor effect.

This point is illustrated in the upper right plot of Fig. \ref{fig:simplecase} where the output fluxes are plotted versus the input flux $\phi_i$. The fact that a small variation of $\phi_i$ makes a large variation in the output fluxes can be seen through the slopes of their variation curves. These slopes are exactly the differential thermal amplification factor $\alpha_l$ introduced in equations (\ref{eq:amplification}). Note that these slopes can go to infinity and even backbend. This happens for example for $T_e=344$ K for $\phi_i\approx23$ W m$^{-2}$. At this point, the only way to increase $\phi_i$ is that the output fluxes exhibit a jump of around 80 W m$^{-2}$ until the system reaches a new equilibrium state. Looking at expression (\ref{eq:phiiex}), we can see that in some conditions the increase in $T_2$ can involve a decrease in the exchange functions $\mathcal{E}_1$ and $\mathcal{E}_2$ which is not compensated by the increasing of $T_2$ so that $\phi_i$ decreases with temperature. The base exhibits here a negative differential thermal resistance \citep{ben-abdallah_near-field_2014} so that there is no stable equilibrium state in a certain temperature range. This is due to the very strong  decreasing of the material emssivity when the material exhibits its phase transition from dielectric to metallic state. We can therefore conclude that the device exhibit two different regimes. In the first one, that we can call the amplification regime, the device is able to make an amplification of the variations of the input flux in the output fluxes. This regime operates as long as the slopes of the curves of the output fluxes variation with the input flux are not backbending. When the curves are backbending, the device behaves like a switch so that we will call this regime the switching regime.  

The lower left plot of Fig. \ref{fig:simplecase} shows the exchange function variations with the base temperature. We note that these variations does not depend on the temperature $T_e$ but only on the temperature $T_2$ which drives the material radiative properties. We note the strong variations of these exchange functions in the temperature range where the VO$_2$ phase transition occur which is not surprising. Note also that the exchange functions  $\mathcal{E}_1$ and $\mathcal{E}_2$ are very close to each other. This is often the case in the transition zone where $T_2$ is close to $T_e$. $\mathcal{E}_1/(\mathcal{E}_1+\mathcal{E}_2)$ and  $\mathcal{E}_2/(\mathcal{E}_1+\mathcal{E}_2)$ that appear in the expression of $\alpha$ are very close to $1/2$ so that the transistor effect is completely dominated by the variations of $\mathcal{E}_e+\mathcal{E}_t$ with the input flux $\phi_i$. In this plot, we note also that the device is in the amplification regime as long as the variations of the exchange function do not exhibit any backbending otherwise it is in the switching regime.

\subsection{Near-field transfer and opaque base}

\begin{figure}
\includegraphics[width=\textwidth]{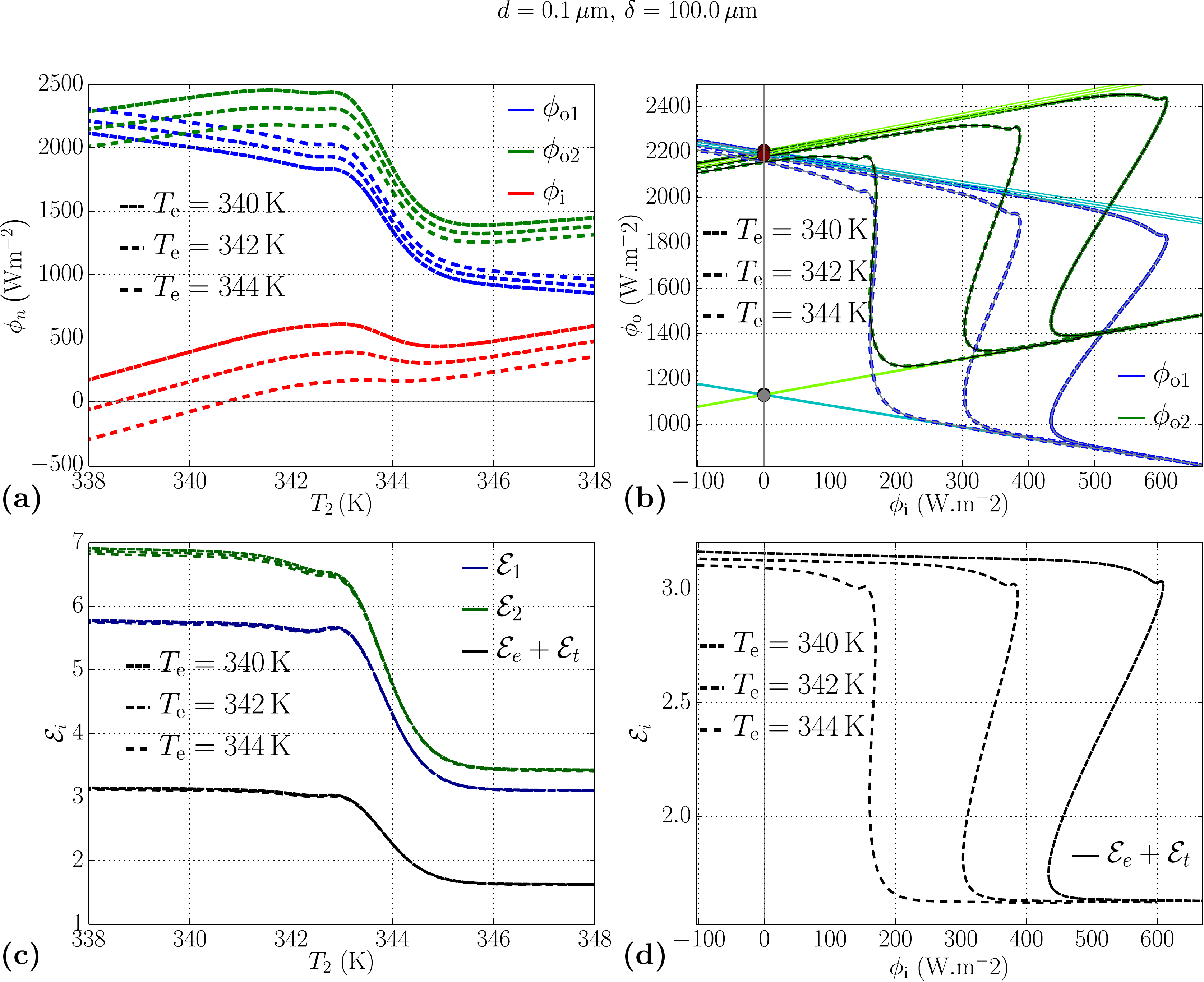}
\captionsetup{justification=centerlast}
\caption{Input and output radiative heat flux densities within the thermal transistor in the opaque ($\delta=100\,\si{\um}$) and near-field ($d=100\,\mathrm{nm}$) case, for three values of the central temperature ($T_{e}=340,342,344\,\mathrm{K}$) and $T_\Delta=280\,\mathrm{K}$.
(a) : ($\phi_{i}$) and output ($\phi_{o1},\phi_{o2}$) fluxes versus the base temperature $T_2$.
(b) : Output fluxes ($\phi_{o1},\phi_{o2}$) versus input flux $\phi_i$.
(c) : Exchange functions ($\mathcal{E}_1$, $\mathcal{E}_2$, $\mathcal{E}_e+\mathcal{E}_t$) versus $T_2$.
(d) : Exchange function $\mathcal{E}_e+\mathcal{E}_t$ versus $\phi_i$.}
\label{fig:nearopaque}
\end{figure}

We now examine a situation where the distance $d$ is small compared to the thermal wavelength and $\delta$ is large compared to the penetration depth in the VO$_2$. Let us take  $d=0.1$ $\si{\um}$ and $\delta=100$ $\si{\um}$.
Here again, we take $T_\Delta$ to 280 K, and we calculate the fluxes and the exchange functions for three values of $T_e$ : 340 K, 342 K and 344 K. The respective values of $T_1$ and $T_3$ are the same than in the preceding case.
The values of the input and output fluxes as well as exchange functions are represented  versus $T_2$ and $\phi_i$ in Figure \ref{fig:nearopaque}, for the three values of $T_{e}$.

We see that the general behavior is similar to the far-field and opaque base case. The main difference is in the amplitude of the fluxes and the exchange functions. In particular we see that the amplitude of the fluxes is one order of magnitude larger than in the far-field case. This is due to the additional heat transfer arising in the near-field due to the presence of evanescent waves that has been adressed in numerous paper. Indeed, SiO$_2$ as well as VO$_2$ in the dielectric phase \cite{van_zwol_phonon_2011} exhibit surface phonon-polaritons in the infrared susceptible to strongly increase near-field heat transfer. When VO$_2$ is in the metallic phase, the polariton disappears in VO$_2$ but is still present in SiO$_2$. The transfer is therefore reduced at higher temperature when the metallic phase appears. The fact that heat transfer is larger in the near-field than in the far field can also be seen when one notices that the exchange functions are larger than 1, meaning that the transfer is larger than the one between two blackbodies. One can also note that in this situation the transistor can be, as in the preceding case in the amplification regime (when $T_e=344$ K) or in the switching regime (when $Te=340$ K or 342 K). This configuration is however more interesting since the heat fluxes involved are more important but of course much more difficult to operate due to the control of the distance of the plates that has to be managed at the nanometric scale.

\subsection{Far-field transfer and semi-transparent base}

The first two preceding situations have actually, in a different way, already been adressed in the past by Ben-abdallah and Biehs \cite{ben-abdallah_near-field_2014} and Joulain et al. \cite{joulain_modulation_2015} where the base is always considered as opaque. However, as already mentioned, in order to have an opaque base, one has to consider a sufficiently thick plate of VO$_2$. The more this plate is thick, the more its thermal inertia (phase transition latent heat of heat capacity) will be important. Even, if the dynamical behavior of the transistor is not the goal of this study, it is important to see if the thermal transistor effect still exists for a thin base and if we can intend to use it for future fast application for thermal amplification or modulation.
 
\begin{figure}
\includegraphics[width=\textwidth]{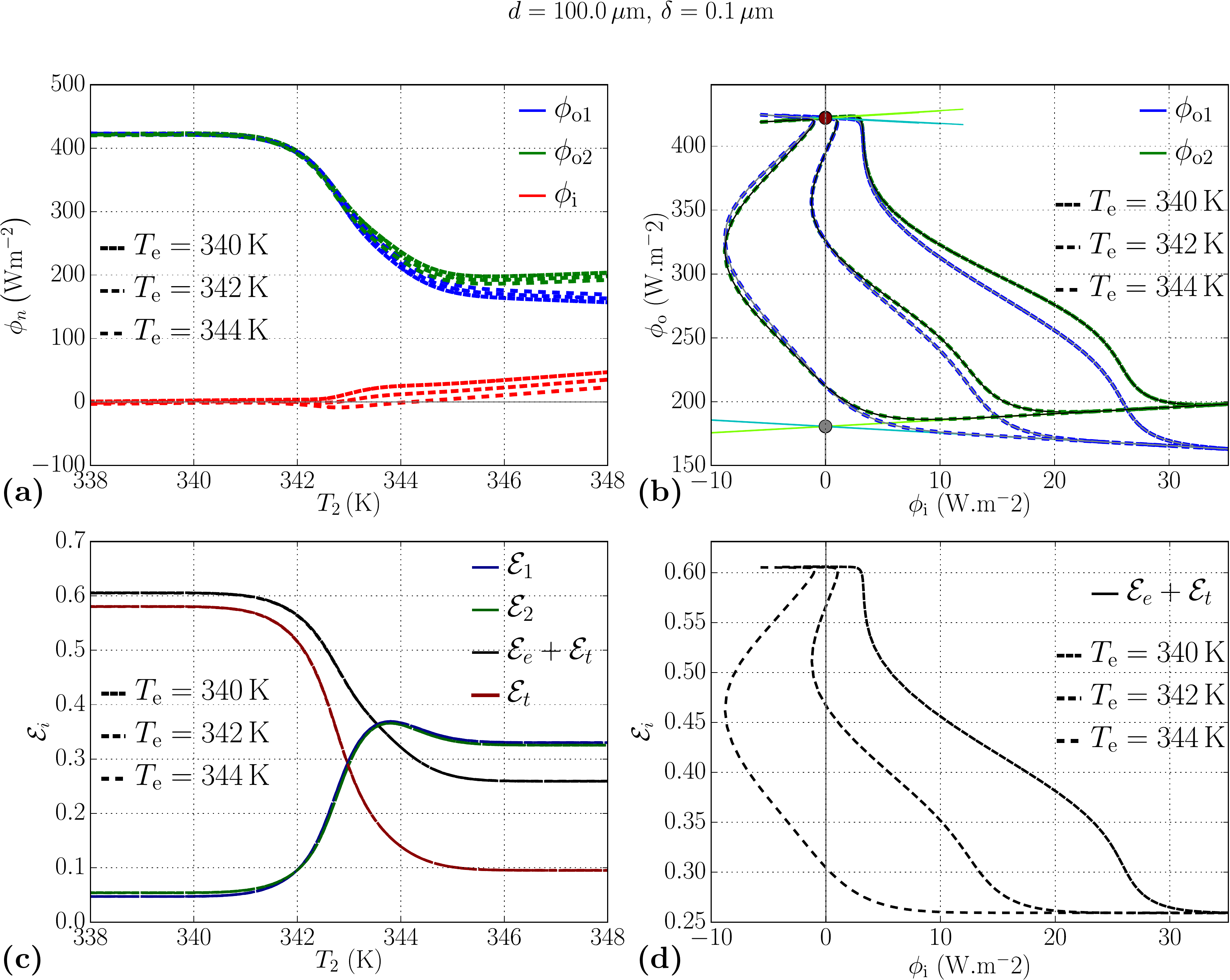}
\captionsetup{justification=centerlast}
\caption{Input and output radiative heat flux densities within the thermal transistor in the semi-transparent ($\delta=100\,\mathrm{nm}$) and far-field ($d=100\,\si{\um}$) case, for three values of the central temperature ($T_{e}=340,342,344\,\mathrm{K}$) and $T_\Delta=280\,\mathrm{K}$.
(a) : ($\phi_{i}$) and output ($\phi_{o1},\phi_{o2}$) fluxes versus the base temperature $T_2$.
(b) : Output fluxes ($\phi_{o1},\phi_{o2}$) versus input flux $\phi_i$.
(c) : Exchange functions ($\mathcal{E}_1$, $\mathcal{E}_2$, $\mathcal{E}_e+\mathcal{E}_t,\mathcal{E}_t$) versus $T_2$.
(d) : Exchange function $\mathcal{E}_e+\mathcal{E}_t$ versus $\phi_i$.}
\label{fig:fartransp}
\end{figure}

We first examine a situation where the distance $d$ is large compared to the thermal wavelength and $\delta$ is small compared to the penetration depth in the VO$_2$. Let us take  $d=100$ $\si{\um}$ and $\delta=100$ nm.
We take $T_\Delta$ to 280 K, and we calculate the fluxes and the exchange functions for three values of $T_e$ : 340 K, 342 K and 344 K. The respective values of $T_1$ and $T_3$ are the same than in the preceding cases.
The values of the input and output fluxes as well as exchange functions are represented  versus $T_2$ and $\phi_i$ in Fig. \ref{fig:fartransp}, for the three values of $T_{e}$.

As we came back to a far-field situation, let us first note, as expected, that heat fluxes amplitudes are of the same order than in the first situation and that the exchange functions remain smaller than 1. An interesting feature can be seen in the plot of the exchange functions variations with the base temperature. One notices that $\mathcal{E}_1$ and $\mathcal{E}_2$ exhibit a maximum. Moreover, contrary to the opaque case, the exchange functions are lower when the VO$_2$ base is in the dielectric case than in the situation where it is in the metallic case. This situation can be explained as follows; when the base is in the dielectric phase, the material is mostly transparent so that it absorbs little of incident radiation. When the temperature increases, the phase transition occurs so that islands of metal appear in the material that become more and more absorbing. The exchange function increases until it reaches a maximum. Indeed, at the end of the phase transition, the material becomes completely metallic and mostly reflective. Absorption only occurs on a small thickness equal to the penetration depth in metal and the exchange function decreases. Another interesting feature is that the evolution of the output fluxes with the input fluxes is different from the two cases we discussed previously. For example, if $T_e=342$ K, one can see that both regime (amplification and switching) are present. Indeed, one observes that the variation curves of the output fluxes first present a backbending (switching) followed by a zone where $\phi_{o1}$ and $\phi_{o2}$ decreases strongly but monotonically with $\phi_i$ (amplification). Depending on the application pursued, one can choose the working point in order to exhibit one or the other property.

\subsection{Near-field transfer and semi-transparent base}
In this last situation, the distance $d$ is small compared to the thermal wavelength and $\delta$ is small compared to the penetration depth in the VO$_2$. We take  $d=100$ nm and $\delta=100$ nm. $T_\Delta$ and the different $T_e$ considered are always the same as before. 

\begin{figure}
\includegraphics[width=\textwidth]{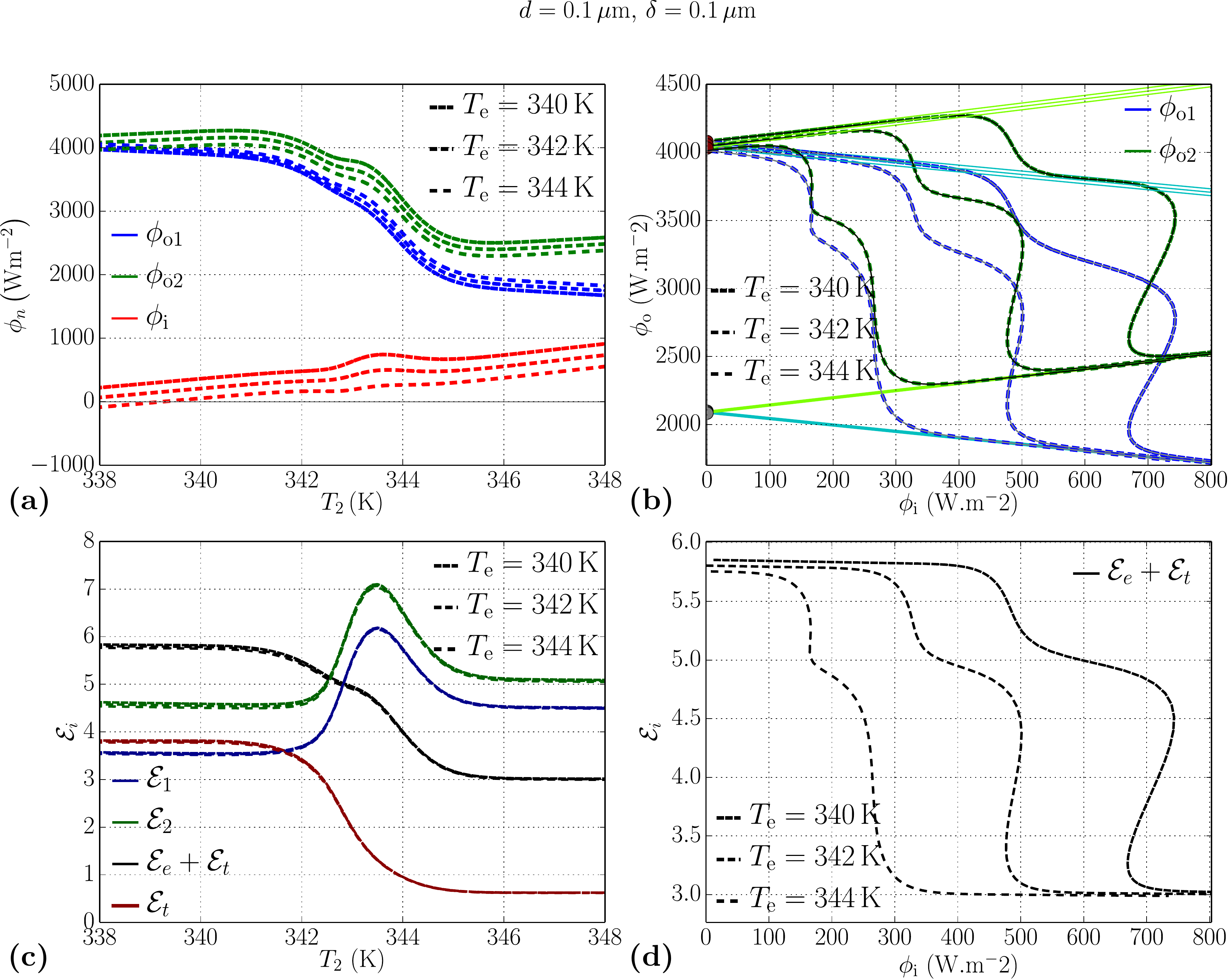}
\captionsetup{justification=centerlast}
\caption{Input and output radiative heat flux densities within the thermal transistor in the semi-transparent ($\delta=100\,\mathrm{nm}$) and near-field ($d=100\,\mathrm{nm}$) case, for three values of the central temperature ($T_{e}=340,342,344\,\mathrm{K}$) and $T_\Delta=280\,\mathrm{K}$.
(a) : ($\phi_{i}$) and output ($\phi_{o1},\phi_{o2}$) fluxes versus the base temperature $T_2$.
(b) : Output fluxes ($\phi_{o1},\phi_{o2}$) versus input flux $\phi_i$.
(c) : Exchange functions ($\mathcal{E}_1$, $\mathcal{E}_2$, $\mathcal{E}_e+\mathcal{E}_t$, $\mathcal{E}_t$) versus $T_2$.
(d) : Exchange function $\mathcal{E}_e+\mathcal{E}_t$ versus $\phi_i$.}
\label{fig:neartransp}
\end{figure}

This case, presented on Fig. \ref{fig:neartransp}, is very similar to the preceding one. It exhibits the behavior we described before concerning the maximum in the exchange functions for a temperature in the middle of the transition phase temperature range. Moreover, the amplitude of the flux is much larger due to the fact that the radiative heat transfer is now in the near-field. This increase is happening again due to the presence of polaritons in both SiO$_2$ and VO$_2$ when this last material is in the dielectric state. Note that transfer is also enhanced due to the coupling of polaritons in the film forming the gate. When the VO$_2$ is in the metallic state, the coupling is less efficient and the output fluxes decrease. In this situation, we also note that both the switching and the amplification regime are present for a same temperature $T_e=344$K. When increasing $\phi_i$, the output fluxes first decrease smoothly (amplification regime) before the curves backbend (switching regime).

\subsection{Modulation efficiency}
\begin{figure}
\begin{center}
\includegraphics[width=\textwidth]{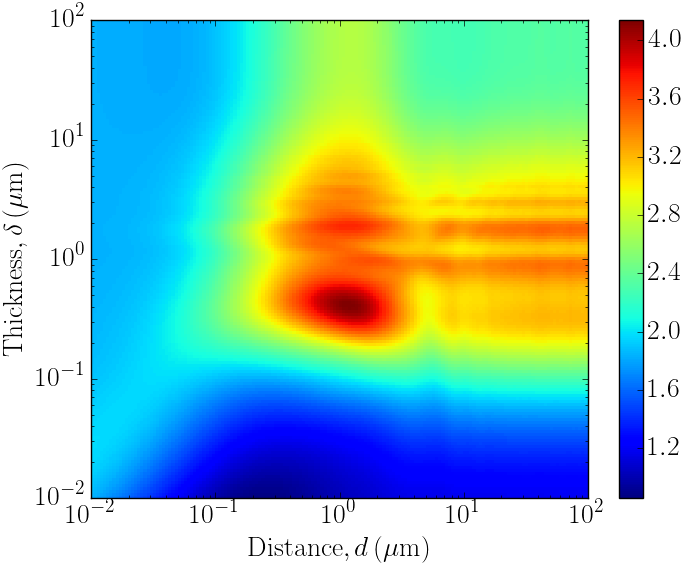}
\caption{Modulation efficiency $ME=\phi_{o}^\mathrm{min}/\phi_{o}^\mathrm{max}$ for different combinations of $(d,\delta)$ and $T_e=342$ K. (The $\phi_o^\mathrm{min}$ values are taken for $T_2=341$ K and the $\phi_o^\mathrm{max}$ values for $T_2=346$ K.}
\label{ratio}
\end{center}
\end{figure}

We now quantify the modulation efficiency of such a device. We have seen that the thermal transistor presented here can exhibit two regimes. However, there is one common point between these regimes : output fluxes are always larger when the base is in the dielectric phase than when it is in the metallic phase. The ratio between this two fluxes determines the modulation efficiency $ME$ as defined in {\it the principle of the radiative thermal transistor} section. Note that this efficiency can also be calculated as the ratio of the exchange functions, so that, as the exchange functions are negligibly varying with $T_e$, the modulation efficiency is negligibly depending on $T_e$ as well, and does not vary according to $T_\Delta$. We calculate this ratio for all combinations of the parameters $(d,\delta)$ and for $T_e=342$ K, the result is presented in Fig. \ref{ratio}.
The modulation efficiency has a maximum of 4.13 for $d_\mathrm{max}\approx1.2\,\si{\um}$ m and $\delta_\mathrm{max}\approx0.4\,\si{\um}$. This point is therefore the one that has to be choose if one wants to make the most important modulation amplitude with this transistor.

\begin{figure}
\begin{center}
\includegraphics[width=8cm]{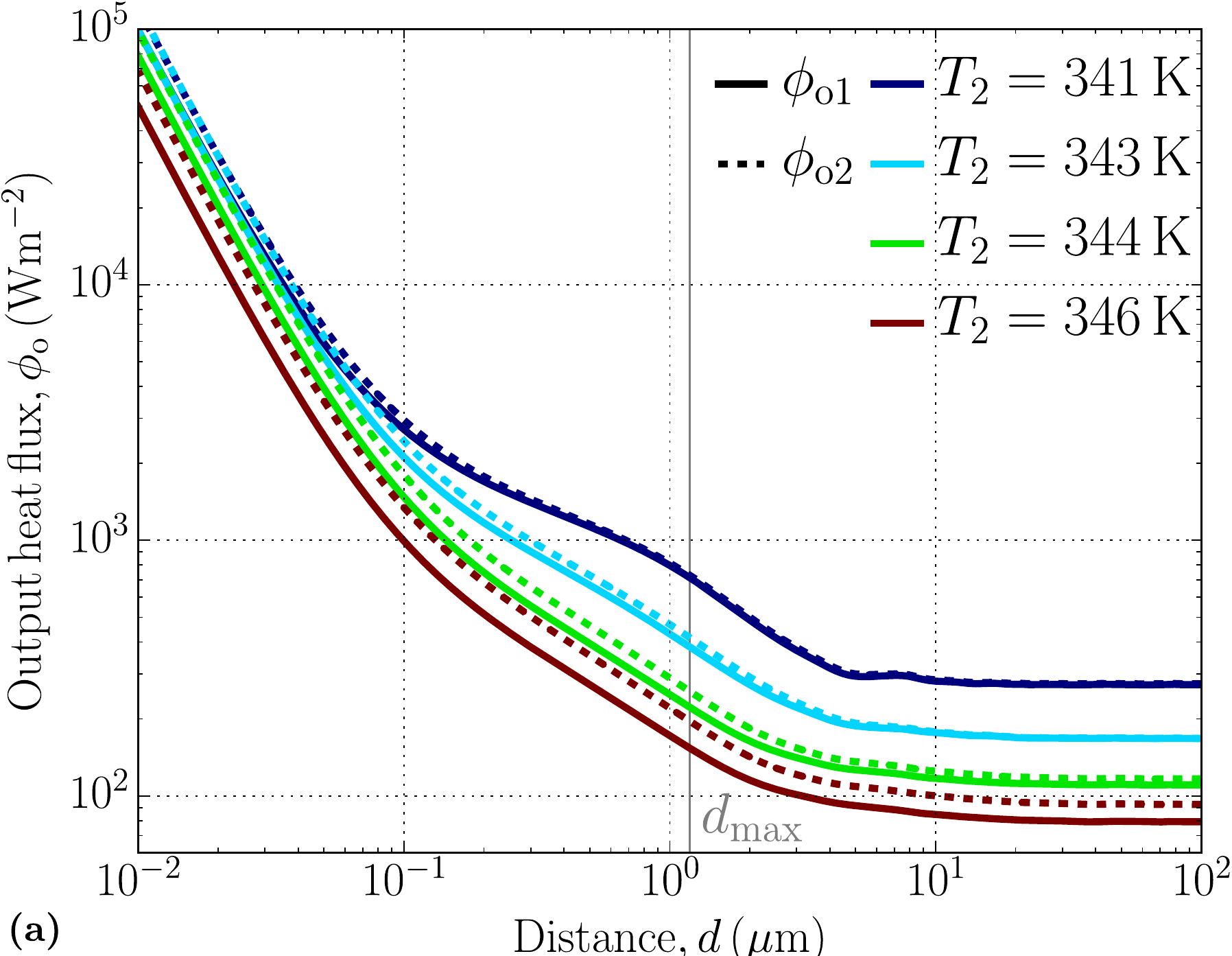}
\includegraphics[width=8cm]{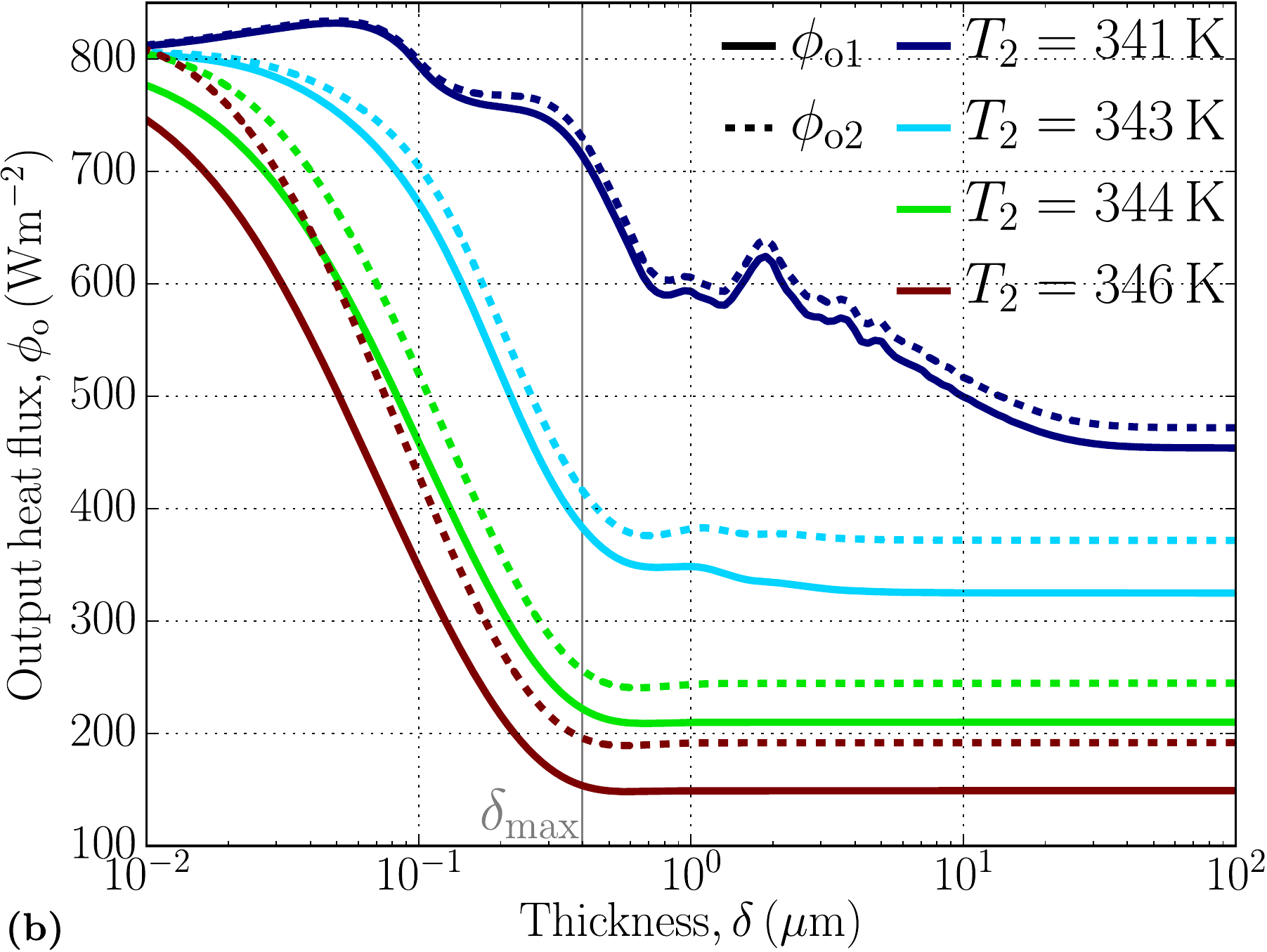}
\caption{(a) Output fluxes versus $d$ for a thickness $\delta$ of 400 nm. (b) Output fluxes versus $\delta$ for a separation distance $d$ of 1.2 $\si{\um}$. These values are calculated with $T_e=342$ K and $T_\Delta=$ 280 K. The VO$_2$ goes from its dielectric state at 341 K to its metallic one at 346 K.}
\label{var_d_delta}
\end{center}
\end{figure}

This maximum can be explained by examining the output fluxes variations with the separation distance $d$ and the thickness $\delta$. Let us begin by the top figure of Fig. \ref{var_d_delta}. It represents the output fluxes versus the separation distance $d$ between the plates. The large distance range corresponds to the far-field whereas the nanometric distance range corresponds to the near-field. In between, there is a transition zone in the micronic zone. When VO$_2$ is in the dielectric phase for the lowest temperature, one notices that the transition zone exhibits a shoulder caracteristic of a polariton \cite{henkel_spatial_2000}. On the contrary, when VO$_2$ is in the metallic state, this shoulder is not present. Therefore, it is in when $d$ is micronic that the ratio between the maximum output flux (when VO$_2$ is a dielectric) and the minimum output flux (when VO$_2$ is a metal) is maximum. The corresponding value $d_\mathrm{max}$ is annotated on Fig. \ref{var_d_delta}(a).
Let us now focus on the variations with $\delta$. When $\delta$ reduces, there is less and less absorption in the film so that the flux increases until it saturates to a value equivalent to a situation where the film would be absent ($\delta\rightarrow 0$). However, when the VO$_2$ base is in the dielectric case, it supports phonon-polaritons that can interact from one side of the film to the other one. At a certain value of the thickness $\delta$ (annotated as $\delta_\mathrm{max}$ on Fig. \ref{var_d_delta}(b)) there exist a resonance in the film where the transfer is enhanced so that the modulation efficiency also reaches a maximum.

\section{Conclusion}
We have studied a radiative thermal transistor made of two plates of SiO$_2$ and one plate of VO$_2$, this last material which exhibits an insulator to metal transition around 340 K. We particularly focused on the influence of the distance $d$ between the elements and thickness $\delta$ of the VO$_2$ plate. We have seen that two regimes can be identified : an amplification regime in which a small change in the input flux generates an amplification on the output fluxes and a switching regime where an input flux applied above a critical value causes a jump to lower values of the output fluxes. We have identified the transistor behavior in the near and the far field as well as in the case of an opaque VO$_2$ base or a transparent one. We have found that the modulation efficiency is maximum for a certain set of the parameters  $d_\mathrm{max}\approx1.2\,\si{\um}$ m and $\delta_\mathrm{max}\approx0.4\,\si{\um}$. In the future, we plan to pursue this work in two directions. Second, we would like first to study the thermal modulation in the dynamical regime taking into account the phase change latent heat and heat capacity of VO$_2$. We would like also to take into account the thermal hysteresis that exists in the VO$_2$ transition phase that for sure plays a very important role on the thermal performance of the radiative transistor..

\begin{acknowledgments}
This work pertains to the French Government Program ''Investissement d'avenir'' (LABEX INTERACTIFS, ANR-11-LABX-0017-01).
\end{acknowledgments}


%

\end{document}